\documentclass{elsart}
\usepackage{ifpdf}
\usepackage{graphicx,amssymb,lineno}
\journal{Journal of Computational Physics}
\ifpdf
\usepackage[%
  pdftitle={Lanczostransport},%
  pdfauthor={Claudio Verdozzi},%
  pdfsubject={The preprint document class elsart},%
  pdfkeywords={instructions for use, elsart, document class},%
  pdfstartview=FitH,%
  bookmarks=true,%
  bookmarksopen=true,%
  breaklinks=true,%
  colorlinks=true,%
  linkcolor=blue,anchorcolor=blue,%
  citecolor=blue,filecolor=blue,%
  menucolor=blue,pagecolor=blue,%
  urlcolor=blue]{hyperref}
\else
\usepackage[%
  breaklinks=true,%
  colorlinks=true,%
  linkcolor=blue,anchorcolor=blue,%
  citecolor=blue,filecolor=blue,%
  menucolor=blue,pagecolor=blue,%
  urlcolor=blue]{hyperref}
\fi

\usepackage{epsfig}
\usepackage{bm}
\usepackage{dcolumn}
\usepackage{amsmath}
\usepackage{graphicx}
\usepackage{latexsym}
\usepackage{amsfonts}
\usepackage{amssymb}

\newcommand{\fff}{\mbox{\scriptsize \boldmath $B$}}
\newcommand{\AB}{{\it ab initio }}

\newcommand{\be}{\begin{equation}}
\newcommand{\ee}{\end{equation}}
\newcommand{\beq}{\begin{eqnarray}}
\newcommand{\eeq}{\end{eqnarray}}

\makeatletter
\def\elsartstyle{%
    \def\normalsize{\@setfontsize\normalsize\@xiipt{14.5}}
    \def\small{\@setfontsize\small\@xipt{13.6}}
    \let\footnotesize=\small
    \def\large{\@setfontsize\large\@xivpt{18}}
    \def\Large{\@setfontsize\Large\@xviipt{22}}
    \skip\@mpfootins = 18\p@ \@plus 2\p@
    \normalsize
}
\@ifundefined{square}{}{\let\Box\square}
\makeatother

\begin{document}
\def\gC{\textbf{C}}
\def\gZ{\textbf{Z}}
\def\gR{\textbf{R}}
\def\gN{\textbf{N}}
\def\bcalS{\textbf{{\cal S}}}
\def\bS{\textbf{\Sigma}}
\def\bG{\textbf{\Gamma}}
\def\bK{\textbf{K}}
\def\bH{\textbf{H}}
\def\bbS{\textbf{S}}
\def\bM{\textbf{M}}
\def\bT{\textbf{T}}
\def\bcalG{\textbf{G}}
\def\bU{\textbf{W}}
\def\bQ{\textbf{B}}
\def\bV{\textbf{V}}
\def\bg{\textbf{g}}
\def\bh{\textbf{h}}
\def\bq{\textbf{q}}
\def\bp{\textbf{p}}
\def\bj{\textbf{j}}
\def\br{\textbf{r}}
\def\cZ{{\cal Z}}
\def\ua{\uparrow}
\def\da{\downarrow}
\def\a{\alpha}
\def\b{\beta}
\def\g{\gamma}
\def\G{\Gamma}
\def\d{\delta}
\def\D{\Delta}
\def\e{\epsilon}
\def\ve{\varepsilon}
\def\z{\zeta}
\def\h{\eta}
\def\th{\theta}
\def\k{\kappa}
\def\l{\lambda}
\def\L{\Lambda}
\def\m{\zeta}
\def\n{\nu}
\def\x{\xi}
\def\X{\Xi}
\def\p{\pi}
\def\P{\Pi}
\def\r{\rho}
\def\s{\gamma}
\def\S{\Sigma}
\def\t{\tau}
\def\f{\phi}
\def\vf{\varphi}
\def\F{\Phi}
\def\c{\chi}
\def\w{\omega}
\def\W{\Omega}
\def\Q{\Psi}
\def\q{\psi}
\def\de{\partial}
\def\inf{\infty}
\def\ra{\rightarrow}
\def\bra{\langle}
\def\ket{\rangle}
\def\rd{{\rm d}}
\begin{frontmatter}
\title{Lanczos-adapted time evolution for open boundary quantum transport}

\author{Claudio Verdozzi \corauthref{cor}}, \ead{cv@teorfys.lu.se}
\author{Carl-Olof Almbladh}\ead{coa@teorfys.lu.se}
\corauth[cor]{Corresponding Author}
\address{Mathematical Physics, Lund Institute of Technology, SE-22100 Lund, Sweden}
\begin{abstract}
We increase the efficiency of a recently proposed time integration scheme for 
time dependent quantum transport by using the Lanczos method
for time evolution. We illustrate our modified scheme
in terms of a simple one dimensional model. 
Our results show that the Lanczos-adapted scheme gives a large
increase in numerical efficiency, and is an advantageous route for numerical
time integration in \AB treatment of open boundary quantum transport phenomena.
\end{abstract}

\begin{keyword}
Quantum transport, Lanczos method, Time evolution
\PACS 72.10. Bg, 76.63.$-$b
\end{keyword}
\end{frontmatter}

\section{Introduction}
\label{intro}

In many physical phenomena, practical limitations hinder a complete knowledge of all the degrees of freedom involved. 
Nanoscience has adopted such apparent shortcoming as its central paradigm, by exploiting the notion of a small system 
coupled to a macroscopic environment. A case in point is represented by quantum transport phenomena, where two (or more)
macroscopic leads are connected to a small central device (quantum constriction).\newline  
Theoretical
approaches to quantum transport can be broadly grouped in two categories, those based on a steady state formulation and
those using a time dependent framework. Another discerning criterion can be 
the type of method used. In this case one can primarily distinguish among \AB  or  model Hamiltonian methods.
Finally, one can also consider a distinction based on the mathematical technique used: nonequilibrium-propagator,
linear-response, wavefunction-scattering, etc. 
Here we consider the time dependent quantum transport (TDQT) approach, 
which permits to follow the system during its time evolution after a bias has been applied. In this way, steady-state, 
transients and a.c. currents can all be considered on equal footing and, in the presence of dissipation, 
history dependence (memory effects) are also accounted for. An early formulation along these lines
was introduced almost three decades ago \cite{Cini80}.\newline
For a quantitative description of
TDQT, as for example required to obtain a theoretical figure of merit of the transient response of a real device, a 
description at the \AB level is certainly required. For this, one can resort
to Time Dependent Density Functional Theory (TDDFT) \cite{rg84,TDDFTbook}. In TDDFT, the TDQT problem is rigorously mapped 
onto a fictitious independent particle problem. A formulation of TDQT within TDDFT has been introduced recently \cite{gstcoa}. The practical applicability 
of the method has also been shown \cite{practical}, 
and the formulation has been extended to include classical nuclear degrees of freedom \cite{cvgstcoa}.\newline 
The purpose of this short communication is to show how the Lanczos algorithm for time evolution \cite{JChemphys}
can be applied to the case of open geometries as those encountered in time dependent quantum transport. 
This is done introducing a modification to the approach given in Ref. \cite{practical}.
After a quick presentation of the Lanczos algorithm, we will review the method in Ref. \cite{practical}.
Then we present our Lanczos-adapted method, and show comparative results for a model
system, followed by some conclusive remarks.
%
%
\section{The Lanczos method}
We briefly summarize the Lanczos method, as given in \cite{JChemphys}. 
A useful comparative study between the Lanczos method and other integration schemes
can be found in \cite{CastroMarques}.
Consider a system described by a TD Hamiltonian $H(t)$. If, for example,  we use the 
mid-point approximation for the time propagator and wish to evolve the system 
in the time interval $(t+\Delta, t)$, we obtain
\begin{equation}
|\Phi_{t+\Delta} \rangle=e^{-iH(t+\Delta/2)\Delta} | \Phi_t\rangle
\end{equation}
where $|\Phi_t\rangle$ is the (known) initial wavefunction.
Consider a finite Lanczos sequence $\{|V_k\rangle\}$, obtained by starting acting
on the  'seed'  $|\Phi_t\rangle\equiv |V_0\rangle $. Using $\{|V_k\rangle\}$ as a 
truncated basis, we get 
\begin{equation}
|\Phi_{t+\Delta}\rangle\approx\sum_{k=0}^{M_L} |V_k\rangle \; \langle V_k| e^{-iH_Lt} |V_0\rangle,\label{apross}
\end{equation}
where $H_L$ is the tridiagonal representation for $H(t+\Delta/2)$ in such a basis.
Inserting a complete set of eigenstates for the truncated space, 
 $H_L|\lambda\rangle=\epsilon_\lambda|\lambda\rangle$,
\begin{equation}
|\Phi_{t+\Delta}\rangle =\sum_{k=0}^{K} |V_k\rangle \; \left[\sum_\lambda  \langle V_k|\lambda\rangle  e^{-i \epsilon_{\lambda} t} \langle\lambda |V_0\rangle\right],
\end{equation}
where $|\Psi_{t+\Delta}\rangle$ is finally expressed in the basis of the original many body Hamiltonian.
The method requires a partial orthogonalization on the fly of the Lanczos basis in order to preserve accuracy along the trajectory.  For a simple estimate of the truncation error in Eq.(\ref{apross}), see the discussion in \cite{JChemphys}.
\section{Boundary Conditions in Time Dependent Quantum Transport} 
An effective and viable strategy to TDQT is to consider large but
finite systems.  Via an initial charge imbalance, a quasi-steady state current
can be established, as clearly shown either in presence of electron-nuclear interactions \cite{Horsfield} 
or when only electrons are considered \cite{DiVentrafinite}. 
A different approach, the one we consider here, is based on an open boundary
formulation of the problem \cite{Cini80, gstcoa}, with a central region connected to two semi-infinite leads \cite{practical}. 
This approach has also been used in a mixed quantum-classical scheme
to deal with electron -phonon systems in quantum transport, where the phonons
are treated as classical fields (Ehrenfest Dynamics, ED)\cite{cvgstcoa}.
It is for this latter approach, which has recently received some attention in the literature,  
that we present a Lanczos-adapted numerical scheme. We will
consider for simplicity the purely electronic case: classical nuclear degrees of freedom 
can be added in a straightforward manner. Finally, a time-dependent embedding 
scheme has been considered very recently also in \cite{Ingles}.
\section{Time Evolution for Quantum Transport}
\label{TEQT}
\noindent We provide here a brief presentation of the open boundary algorithm of Ref.\cite{practical}.
No attempt of completeness is made and we refer to the original paper for a detailed
derivation. In the following, we present the main formulas in the case of a strictly 1D system (i.e. the leads have no translational invariance in the transverse direction), to provide the background needed to introduce our Lanczos-adapted scheme. The Hamiltonian we consider is  $\bH^{tot}(t)=\bH_{el}+\bU(t)$, where $\bU(t)$ is the external perturbation. 
In a TDDFT approach, the initial, ground state is a single Slater determinant $|\Q_{g}\ket$. It is useful to divide the (1D) space into three regions. With $s$ a the site label, we have the region $L$ (corresponding to the left lead,  with $s\le-(M+2)$ ), the central region C ( with  $|s|\le M+1$ , i.e the device region contains $2M+3$ sites), and the region $R$ (corresponding to right lead,  with  $s\ge(M+2) )$.  The general structure of any bound, extended or resonant one particle eigenstate $\q$ in the
Slater determinant $|\Q_{g}\ket$ can be written as 
\be
\q(s)
=\left\{
\begin{array}{ll}
L_{+}e^{-ik_{l} s}+L_{-}e^{ik_{l} s} & \;\; s\leq-M-2 \\
\q(s)  & \;\; |s| \leq M+1 \\
R_{+}e^{ik_{r } s}+R_{-}e^{-ik_{r} s} & \;\; s\geq M+2 \\
\end{array}
\right.,
\label{gsops}
\ee
To describe quantum transport, one needs to evolve in time the 
ground state configuration $|\Q_{g}\ket$, i.e. each one of the single particle eigenstates $\q$ above. 
Introducing the projection operators ${\bf P}_{L,C.R}$ (for example,  ${\bf P}_{L}=\sum_{s\in L}|s\ket\bra s|$), we can write ($\b=L,C,R$), for the generic single particle state,
\begin{equation}
|\q\ket=\sum_{\b}|\q_{\b}\ket,\quad |\q_{\b}\ket={\bf P}_{\b}|\q\ket.\label{short2}
\end{equation}
In the same way, we can project the Hamiltonian in the different regions
\begin{equation}
\bH=\sum_{\b\b'}\bH_{\b\b'},\quad \bH_{\b\b'}\equiv 
{\bf P}_{\b}\bH{\bf P}_{\b'}.\label{short1}
\end{equation}
Separating the contribution from the leads in $\bU$, the set of one-particle equations becomes
\be
i\frac{\rd}{\rd t}|\q(t)\ket=
\left[\bH(t)+\bU_{leads}(t)\right]|\q(t)\ket,
\ee
with $\bH(t)=\bH_{\rm el}+\bU_{CC}(t)$,
where  $\bH_{\rm el}$ 
is the electron one particle Hamiltonian and $\bU_{CC}(t)$ is the external potential projected in the central 
region $C$.
Assuming metallic electrodes \cite{practical},
\be
\bU_{leads}(t)
=\left\{
\begin{array}{ll}
\d_{s,s'}W_{L}(t) & \;\; s \leq -M-2 \\
0 & \;\; s \leq |M+1| \\
\d_{s,s'}W_{R}(t) & \;\; s \geq M+2 \\
\end{array}
\right..
\ee
In the numerical time propagation, the time is discretized:  $t_{m}=2m \d$ , where 
$\d$ is the timestep, $m$ is an integer, and the explicit prefactor $2$ is introduced
for convenience in the formulas. In  \cite{practical}, the one-particle eigenstates are propagated from $t_{m}$ to $t_{m+1}$ using a generalised Crank-Nicholson scheme. 
For the time evolution of each one of the one-particle states in 
$|\Q_{g}\ket$, one gets \cite{practical}
\be
({\bf 1}+i\d \bH^{(m)})\frac{{\bf 1}+i\frac{\d}{2}\bU_{ 
leads}^{(m)}}{{\bf 1}-i\frac{\d}{2}\bU_{leads}^{(m)}}|\q^{(m+1)}\ket=
({\bf 1}-i\d \bH^{(m)})\frac{{\bf 1}-i\frac{\d}{2}\bU_{ 
leads}^{(m)}}{{\bf 1}+i\frac{\d}{2}\bU_{leads}^{(m)}}|\q^{(m)}\ket,\;\;\;
\label{cn1}
\ee
where $|\q^{m}\rangle\equiv|\q(t_m)\rangle$ and
\begin{eqnarray}
\bH^{(m)}&=&\bH_{\rm el}+\frac{1}{2}\left[\bU_{CC}(t_{m+1})+\bU_{CC}(t_{m})\right]\\
\bU_{leads}^{(m)}&=&\frac{1}{2}[\bU_{leads}(t_{m+1})+\bU_{leads}(t_{m})].
\end{eqnarray}
\subsection{Propagation of One-Particle Eigenstates}
Using Eqs.(\ref{short2},\ref{short1}), and after some algebra, the closed equation for the time-evolution in the central region is
\beq
|\q_{C}^{(n+1)}\ket=
\frac{{\bf 1}_{C}-i\d \bH^{(n)}_{\rm eff}}
{{\bf 1}_{C}+i\d \bH^{(n)}_{\rm eff}}|\q_{C}^{(n)}\ket
-2i\d\sum_{\a=L,R}\frac{\Omega_{\a}^{(n)}}{w_{\a}^{(n)}}
\left(
|\s_{\a}^{(n)}\ket+|\m_{\a}^{(n)}\ket
\right),
\label{central}
\eeq
where
\begin{eqnarray}
w_{\a}^{(n)}=\frac{1-i\frac{\d}{2}W_{\a}^{(n)}}{1+i\frac{\d}{2}W_{\a}^{(n)}},\\
\Omega_{\a}^{(n)}=\prod_{j=0}^{n}[w_{\a}^{(j)}]^{2},
\end{eqnarray}
and
\begin{eqnarray}
\bH^{(n)}_{\rm 
eff}=\bH^{(n)}_{CC}-i\d\sum_{\a=L,R}\bH_{C\a}\frac{1}{{\bf 1}_{\a}+i\d 
\bH_{\a\a}}\bH_{\a C}
=\bH^{(n)}_{CC}-i\d\sum_{\a=L,R}\bQ_{\a}^{(0)}.\label{Hcentral}
\end{eqnarray}
The $\bQ_{\a}^{(0)}$ matrices have only one non-zero element,
\be
\left[\bQ_{\a}^{(0)}\right]_{s,s'}=b^{(0)}
\left\{
\begin{array}{ll}
\d_{s,-M-1}\d_{s',-M-1} & \quad \a=L \\
\d_{s,M+1}\d_{s,M+1} & \quad \a=R
\end{array}
\right.,
\label{qmm}
\ee
with $b^{(0)}=\frac{-1+\sqrt{1+4\d^{2}V^{2}}}{2\d^{2}}$ and $V$ the hopping parameter
in the leads. 
The expression for the source state $|\s_{\a}^{(n)}\ket$ and the memory 
state
$|\m_{\a}^{(n)}\ket$ are \cite{practical}:
\begin{eqnarray}
|\m_{\a}^{(n)}\ket&=&Z_{\a}^{(n)}
\frac{1}{{\bf 1}_{C}+i\d \bH^{(n)}_{\rm eff}}|u_{\a}\ket,\\
|\s_{\a}^{(n)}\ket&=&G_{\a}^{(n)}
\frac{1}{{\bf 1}_{C}+i\d \bH^{(n)}_{\rm eff}}|u_{\a}\ket
\label{efss}
\end{eqnarray}
where $|u_{\a}\ket$ is a unit vector such that
\be
\bra s|u_{\a}\ket=\left\{
\begin{array}{ll}
\d_{s,-M-1} & \quad \a=L \\
\d_{s,M+1} & \quad \a=R 
\end{array}
\right..
\ee
The scalar quantities $Z_{\a}^{(n)}$ and $G_{\a}^{(n)}$, $\a=L,R$ are given by
\begin{eqnarray}
Z_{\a}^{(n)}&=&\frac{\d}{2i}
\sum_{j=0}^{n-1}
\frac{w_{\a}^{(j)}}{\Omega_{\a}^{(j)}}
\left(b^{(n-j)}+b^{(n-j-1)}\right)
\left(\bra u_{\a}|\q_{C}^{(j+1)}\ket+\bra u_{\a}|\q_{C}^{(j)}\ket\right),\label{open1}\\
G_{\a}^{(n)}&=&
\left(
\a_{+}e^{iz_{\a}(M+2)}+\a_{-}e^{-iz_{\a}(M+2)}
\right)
V\frac{\left(1-2i\d\cos\left(z_{\a}\right)\right)^{n}}
{\left(1-2i\d\cos\left(z_{\a}\right)\right)^{n+1}}
\label{sst} \\ 
&+&
\left(
\a_{+}e^{iz_{\a}(M+1)}+\a_{-}e^{-iz_{\a}(M+1)}
\right)
\times
i\d\sum_{j=0}^{n}
\frac{\left(1-2i\d\cos\left(z_{\a}\right)\right)^{n-j}}
{\left(1-2i\d\cos\left(z_{\a}\right)\right)^{n+1-j}}
\left(b^{(j)}+b^{(j+1)}\right)\nonumber
\end{eqnarray}
and $z_{\a}=k_l$ for $\a=L$ while $z_{\a}=k_r$ for $\a=R$.
For $n>2$, the quantities $b^{(n)}$ in the Eqs.(\ref{open1},\ref{sst})
are obtained by recursion:
\beq
b^{(n)}&=&
\frac{b^{(1)}b^{(n-1)}}{b^{(0)}}-\d^{2}\frac{b^{(0)}b^{(n-2)}}{1+2\d^{2}b^{(0)}}
\label{qm} \\ 
&-&\d^{2}
\sum_{j=1}^{n-1}\frac{\left(
b^{(j)}+b^{(j-1)}+b^{(j-2)}\right)
b^{(n-2-j)}}{1+2\d^{2}b^{(0)}}
\nonumber
\eeq
and  
$b^{(n<0)}=0, b^{(1)}=\frac{1-2\d^{2}b^{(0)}}{1+2\d^{2}b^{(0)}}b^{(0)}$ and $b^{(0)}$ the same as in
Eq.(\ref{qmm}).
\section{Lanczos-adapted algorithm}
The basic idea behind the algorithm illustrated in the previous Section
is to discretize the time axis via the Crank-Nicholson algorithm {\em before} performing
the partitioning in L, C, R regions \cite{practical}. One could devise doing the same
for the Lanczos algorithm; however, noncommuting parts of the Hamiltonian would appear in
the exponent this time, rendering formal manipulations more involved. Here, we consider a simple shortcut that, while improving the numerical efficiency of the algorithm in \cite{practical}, has the same
degree of accuracy( i.e. it is second order in $\delta$) but avoids working with the
Lanczos  scheme before the partitioning.
Looking at Eq.(\ref{central}), we notice that the explicit action of $\bH^{(n)}_{\rm eff}$
occurs in two specific terms:
\begin{eqnarray}
|\chi_1\rangle&=&\frac{{\bf 1}_{C}-i\d \bH^{(n)}_{\rm eff}}
{{\bf 1}_{C}+i\d \bH^{(n)}_{\rm eff}}|\q_{C}^{(n)}\ket\label{numb1}\\
|\chi_2\rangle&=&\frac{1}{{\bf 1}_{C}+i\d \bH^{(n)}_{\rm eff}}|u_{\a}\label{numb2}\ket
\end{eqnarray}
where $|\chi_1\rangle$ is the contribution to $|\q_C^{(n+1)}\rangle$ from the central region, and  $|\chi_2\rangle$ enters the expressions for the source and memory states.
For $|\chi_1\rangle$, since $\delta\rightarrow 0$, one can write, up to order two in $\delta$
\be
|\chi_1\rangle=\frac{{\bf 1}_{C}-i\d \bH^{(n)}_{\rm eff}}
{{\bf 1}_{C}+i\d \bH^{(n)}_{\rm eff}}|\q_{C}^{(n)}\ket\approx e^{-2i\delta \bH^{(n)}_{\rm eff}}|\q_{C}^{(n)}\ket.
\ee
For the case of $|\chi_2\rangle$, we define the following quantities:
\beq
\Delta_\pm=\frac{1\pm \sqrt{3}}{2}\delta
\eeq
which permit to rewrite $|\chi_2\rangle$ as
\begin{eqnarray}
|\chi_2\rangle&=&\left[-1+e^{-i\bH^{(n)}_{\rm eff}\Delta _+}
+e^{-i\bH^{(n)}_{\rm eff}\Delta _-}\right] |u_{\a}\ket +O(\delta^3)\label{nice}
\end{eqnarray}
If necessary, one can go to higher orders, by imposing that $(1+\delta x)^{-1}=A+\sum_k e^{a_k \delta x}$
and finding the coefficients $A, \{a_k\}$ by comparison of the two expressions order by order in $\delta$
(in general, the $ \{a_k\}$ will be complex). We note that the same Lanczos sequence of 
basis vectors is required for both exponentials in Eq.(\ref{nice}).\newline
All terms which appear in the propagation scheme \cite{practical} and that involve  ${\bH^{(n)}_{\rm eff}}$, have been re-expressed in terms of exponentials, so that Lanczos propagation can be
used; finally, since ${\bH^{(n)}_{\rm eff}}$ is complex, Eq.(\ref{Hcentral}), it is convenient to split the exponentials;  for small
$\delta$, 
\begin{eqnarray}
e^{-2i\delta \bH^{(n)}_{\rm eff}}&\approx& e^{-\delta^2\sum_\alpha\fff^{(0)}_{\alpha}}
e^{-2i\delta \bH^{(n)}_{CC}}
e^{-\delta^2\sum_\alpha \fff^{(0)}_{\alpha}}\label{split}\\
e^{-i\Delta_\pm \bH^{(n)}_{\rm eff}}&\approx&
e^{-\frac{\delta}{2} \Delta_\pm \sum_\alpha \fff^{(0)}_{\alpha}}
e^{-i\Delta_\pm  \bH^{(n)}_{CC}}
e^{- \frac{ \delta}{2} \Delta_\pm \sum_\alpha \fff^{(0)}_{\alpha}}\nonumber\\\label{split2nd}
\end{eqnarray}
For the 1D case, the advantage is immediate:  the $\bQ_{\a}^{(0)}$ in Eq.(\ref{qmm})
have only one non-vanishing entry and
the outer exponentials in Eq.(\ref{split},\ref{split2nd}) reduce to scalars (here, we do not address the 3D case; however, we expect that the splitting will still provide a simplification).
\section{A Numerical Example}
\begin{figure}[tbp]
\includegraphics*[width=1.0\textwidth]{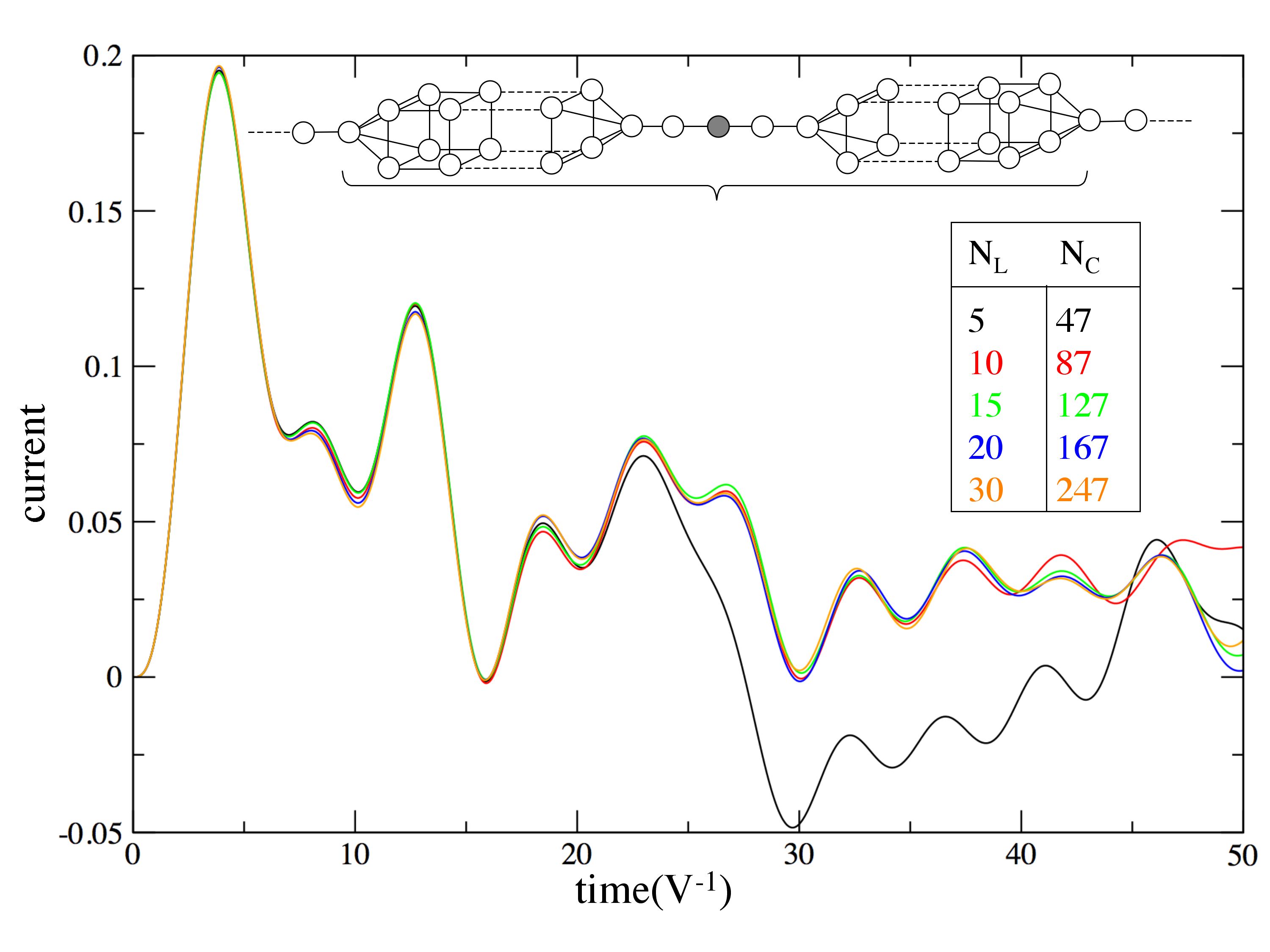}
\caption{Results for the current for different numbers of sites in the central region.
for a bias $U_L=0.5|V|$.  Color coding is specified in the top right inset panel. }
\label{fig1}
\end{figure}

The advantage of the modified scheme presented here is the possibility of manipulating very 
efficiently exponentials (via the Lanczos scheme), thus being
able to deal with larger scale problems in a faster way. 
We have performed some tests for a simple spinless model of QT, namely 
a 3D central region $C$ connected to two semi-infinite 1D metallic leads, in the half-filling regime.
The leads are described by a nearest neighbour, tight binding Hamiltonian (nnTBH) with 
hopping parameter $V=-1$. The central region, as shown in Fig.\ref{fig1}, is 
made of a short central chain of five sites connected with two identical clusters.
Such clusters are composed by periodically repeated layers, each layer containing
four atoms arranged in a square. For technical reasons, the rightmost (leftmost) of 
the left (right) lead is also included in $C$. 
The single particle Hamiltonian in C is also a nnTBH, where $V_C=-0.3$.
The number of $N_C$ sites in the
central region is $N_C=2 N_L+7$, and we vary it by changing $N_L$. 
If we increase $N_L$, we can think of our system as a five-site chain
connected to finite, but progressively longer 3D leads (the latter
are in turn connected to the 1D, truly semi-infinite leads). In our QT simulations, 
there will be a transient, but increasingly longer, time interval before the truly
1D nature of the electron reservoirs will manifest.  
We have analysed the current $j=2 V \sum_k^{occ}Im\left[\psi_k(R)\psi^*_k(R+1)\right]$
at the central site $R=0$ (the grey-shaded circle in Fig.\ref{fig1}) 
as a function of $N_L$. 
For any fixed time $\bar{t}$, on increasing $N_L$ 
the current $j$ converges to a specific value; deviations from the converged
value occur at longer times for greater values of $N_L$, because the 1D nature
of the real reservoirs enters at later stages for longer clusters.
To assess the efficiency of our modified scheme,  
we have  calculated the currents of Fig.\ref{fig1} for different $N_L$
in two ways (which differ on how Eqs.(\ref{numb1}, \ref{numb2})
are computed). Namely, we i) used standard LAPACK routines to compute the inverse of the operator ${\bf 1}_{C}+i\d \bH^{(n)}_{\rm eff}$,
and ii ) used the Lanczos-adapted scheme introduced here. 
We note that, in analogy to  \cite{pumping}, 
another way to manipulate Eq.(\ref{numb1}) is to iii)
solve a linear system, after recasting Eq.(\ref{numb1}) as 
\beq
({\bf 1}_{C}+i\d \bH^{(n)}_{\rm eff} |\chi_1\rangle =({\bf 1}_{C}-i\d \bH^{(n)}_{\rm eff})|\q_{C}^{(n)}\ket. \label{syslin}
\eeq
Such linear system is to be solved for each single particle state, and this is expected
to become computationally unfavourable (unless the
Hamiltonian has a special structure such as band-diagonal, sparse, etc.)
when the number of single particle states 
in the Slater determinant and/or the size of central region become large.
On the other hand, the operator in Eq.(\ref{numb1}) is state-independent,
and the inversion can be performed before entering the loop for the 
single particle states in the Slater determinant.
Accordingly, we did not consider iii) in our numerical comparisons.
In all calculations we used 
a timestep $\delta=0.01|V|^{-1}$, with $N_t$=5000 timesteps. 
For the short iterated Lanczos scheme, we used $6$ iterations/timestep.
Results for the execution times,
as a function of $N_L$  are shown in Table \ref{table1}.
We see that on increasing $N_L$,  the Lanczos adapted scheme becomes
significantly more efficient than i). 
We expect this to be a general
trend: for genuine 3D systems/leads, the advantage of a Lanczos-adapted time evolution
should then become even more significant. At the same time, the actual figures of
relative numerical efficiency between i) and ii) in Table \ref{table1} should be considered only
as indicative, since we have not performed a careful optimization 
of the Lanczos-adapted algorithm/code (an optimized code could further
improve the numerical performance).
\begin{table}
\caption{\label{table1}Comparison between different schemes of 
numerical integration for the algorithm of Ref. \cite{practical} for
central regions of different size. Execution times are in arbitrary
units.}
\begin{tabular}{lccccr}
$N_L$ & 5& 10&15&20&30\\
\hline
Inversion & 1.00 & 1.51 & 2.14 & 3.30 & 14.94 \\
Lanczos  & 1.55 & 2.02 & 2.50 & 2.99 &4.02 \\
\end{tabular}
\end{table}
\section{Conclusions}
In this short note, we have described a simple way to increase the numerical efficiency 
of a recently proposed algorithm for time dependent quantum transport.
We tested the efficiency of the proposed scheme in terms of a model system.
While our modifications to the original algorithm are rather simple, we expect
that the practical advantage of such modifications to be significant,
since future time dependent \AB calculations for quantum transport in realistic
structures are expected to involve sizeable active regions, i.e. large configuration spaces
and large scale calculations. This work was supported by  EU 6th framework Network of Excellence
NANOQUANTA (NMP4-CT-2004-500198).

\end{document}